# The conformational evolution of elongated polymer solutions tailors the polarization of light-emission from organic nanofibers

*Andrea Camposeo[1,‡,*], Israel Greenfeld[2,‡,*], Francesco Tantussi[3,4], Maria Moffa[1], Francesco Fuso[3,4], Maria Allegrini[3,4], Eyal Zussman[2], Dario Pisignano[1,5,*]*

[1]National Nanotechnology Laboratory of Istituto Nanoscienze-CNR, via Arnesano, I-73100 Lecce (Italy)

[2]Department of Mechanical Engineering, Technion – Israel Institute of Technology, Haifa 32000, Israel

[3] Dipartimento di Fisica "Enrico Fermi" and CNISM, Università di Pisa, Largo Bruno Pontecorvo 3, I-56127 Pisa (Italy)

[4] Istituto Nazionale di Ottica INO-CNR, Sezione di Pisa, I-56127 Pisa (Italy)

[5]Dipartimento di Matematica e Fisica "Ennio De Giorgi", Università del Salento, via Arnesano I-73100 Lecce, (Italy)

ABSTRACT. Polymer fibers are currently exploited in tremendously important technologies. Their innovative properties are mainly determined by the behavior of the polymer macromolecules under the elongation induced by external mechanical or electrostatic forces, characterizing the fiber drawing process. Although enhanced physical properties were observed in polymer fibers produced under strong stretching conditions, studies of the process-induced nanoscale organization of the polymer molecules are not available, and most of fiber properties are still obtained on an empirical basis. Here we reveal the orientational properties of semiflexible polymers in electrospun nanofibers, which allow the polarization properties of active fibers to be finely controlled. Modeling and simulations of the conformational evolution of the polymer chains during electrostatic elongation of semidilute solutions demonstrate that the molecules stretch almost fully within less than 1 mm from jet start, increasing polymer axial orientation at the jet center. The nanoscale mapping of the local dichroism of individual fibers by polarized near-field optical microscopy unveils for the first time the presence of an internal spatial variation of the molecular order, namely the presence of a core with axially aligned molecules and a sheath with almost radially oriented molecules. These results allow important and specific fiber properties to be manipulated and tailored, as here demonstrated for the polarization of emitted light.

INTRODUCTION

Fibers[1-4] are typically formed upon the solidification of a tiny filament drawn from a viscous polymer solution or melt,[5-7] whose thinning follows a very complex dynamics.[8-10] Understanding how polymer chains modify their conformation at nanoscale, and to what extent they keep their configuration in solid nanostructures, is fundamental for many applications and for controlling the resulting physical properties of fibers.[11,12] For example, polymers are typically considered bad thermal conductors, but aligning their chains in 1-dimensional (1D) nanostructures allows their thermal conductivity to be improved approaching the single-molecule limit (about 350 W $m^{-1}$ $K^{-1}$ for polyethylene).[5] Similarly, charge mobilities ($\mu$) in organic semiconductor films are typically low (most often < $10^{-1}$ V $cm^{-2}$ $s^{-1}$), although in single $\pi$-conjugated polymer chains $\mu$ can be of the order of hundreds of V $cm^{-2}$ $s^{-1}$.[13] In the bulk, the disordered supramolecular assembly limits the charge mobility, whereas 1D nanostructures show an increase of 1 to 3 orders of magnitude of $\mu$.[14,15] The alignment of $\pi$-conjugated molecules is also effective to improve the amplification of light by stimulated emission,[16] the macroscopic quantum spatial coherence of the exciton state,[17] and the polarization of emitted light. In general, stretching a semidilute polymer solution by an electrostatic field is very effective to prime the formation of fibers, potentially resulting in a structure mostly composed of ordered and aligned chains.[18-20] Little is known however about the nanoscale features induced by elongational dynamics, and about how these features can be exploited to tailor and control macroscale properties of solid nanostructures.

In this paper, we employ the unique features of scanning near-field optical microscopy (SNOM)[21-23] to investigate at nanoscale polymer fibers produced by electrospinning. Absorption measurements with nm-spatial resolution and polarization modulation provide insight into the nanoscale variation of molecular alignment, evidencing an unexpected change from axial to radial molecular orientation upon moving from the fiber axis to its surface. The formation of such complex structures occurs close to the polymer jet start, as demonstrated by modeling the evolution of the conformation of the polymer chains network.

EXPERIMENTAL

*Conjugated polymer nanofibers*. Fibers are produced by electrospinning a solution (70-200 μM polymer) of poly[2-methoxy-5-(2-ethylhexyl-oxy)-1,4-phenylene-vinylene] (MEH-PPV) (molecular weight 380,000 g/mol, American Dye Source Inc). Sprayed films of micro-beads and micro-fibers are obtained at concentrations > 200 μM. The polymer is dissolved in a 1:4 (weight:weight) mixture of dimethylsulfoxide (DMSO) and tetrahydrofuran (THF). The electrospinning system consists of a microprocessor dual drive syringe pump (33 Dual Syringe Pump, Harvard Apparatus Inc.), feeding the polymer solution through the metallic needle at constant rate (10 μL/min). A 11 kV bias is applied between the needle and a metallic collector (needle-collector distance 6 cm), made of two Al stripes positioned at a mutual distance of 2 cm. The MEH-PPV nanofibers are collected on a 1×1 cm$^2$ quartz substrate for optical investigation. Arrays of uniaxially aligned nanofibers are also produced by using a rotating collector (4000 rpm, corresponding to a linear velocity of 30 m/s at the disk edge) for emission polarization measurements, featuring similar morphology and optical properties as samples deposited on the Al stripes. The fiber morphology is investigated by scanning electron microscopy (SEM) using a Nova NanoSEM 450 system (FEI), with an acceleration voltage of 5-10 kV.

*Polarized emission*. Optical images of the fibers are obtained by confocal microscopy, using an inverted microscope (Eclipse Ti, Nikon) equipped with a confocal laser scanning head (A1R MP,

Nikon). An Ar$^+$ ion laser ($\lambda_{exc}$=488 nm) excites the fibers through an oil immersion objective with numerical aperture, $N.A.$ =1.4. The intensity of the light transmitted through the sample, measured by a photomultiplier, is recorded synchronously to the confocal acquisition of the laser-excited fluorescence. The polarization of the emission of individual nanofibers at different polymer concentrations is characterized by a micro-photoluminescence system, composed by a diode laser excitation source ($\lambda$=405 nm) coupled to an inverted microscope (IX71, Olympus). The laser beam propagates perpendicular to the substrate on which the fiber is deposited, and it is focused on the sample through a 20× objective ($N.A.$=0.5, spot size 30 μm). Furthermore, the fiber is positioned with its longitudinal axis almost parallel to the incident laser polarization. The PL emitted by individual nanofibers is collected along the direction perpendicular to the substrate, by means of an optical fiber, and dispersed in a monochomator (USB 4000, Ocean Optics). The polarization of the emission is analyzed by a polarization filter mounted on a rotating stage and positioned between the emitting MEH-PPV nanofiber and the collecting optics. The system response is precisely analyzed in order to avoid artifacts due to the collection and measurement apparatus.

*SNOM*. A polarization-modulation near field microscopy system is used to analyze the linear dichroism of tens of individual MEH-PPV fibers. The SNOM system, operating in emission mode, excites samples in the optical near-field of a tapered optical fiber probe (Nanonics), with a nominal aperture of 50 nm, delivering a near-field power up to the tens of nW range ($\lambda$=473 nm). The signal transmitted by the samples is collected by an aspherical lens ($N.A.$=0.55, diameter of 13 mm) and sent onto a photomultiplier. The polarization modulation relies on a photoelastic modulator (PEM-100, Hinds Instruments), behaving as a waveplate with periodically modulated retardation. The modulator is followed by a $\lambda$/4 waveplate and the whole system is conceived in order to send into the optical fiber probe radiation linearly polarized along a direction periodically oscillating in the transverse plane. The photomultiplier signal is split and sent into two different digital dual lock-in amplifiers (Stanford Research SR830DSP). The first one, referenced to the polarization modulator frequency, $f$, provides with an output (hereafter called AC) representative of the sample response to polarized radiation, whereas the second lock-in, referenced to a slow modulation frequency $f'$ ($f/f' > 10$) of the laser amplitude, is used to determine the optical transmission averaged over all polarization states (DC output).

The dichroic ratio of sample, $\gamma = \dfrac{I_{//} - I_{\perp}}{I_{//} + I_{\perp}}$, where $I_{//}$ and $I_{\perp}$ are the transmitted intensity for polarization aligned along two mutually orthogonal directions, respectively, is quantitatively evaluated from the ratio AC/DC. This requires to model the behavior of the whole optical chain and to account for the residual optical activity of its components, including the optical fiber probe (see Supporting Information). Indeed, reference measurements performed on bare substrates provide a dichroic ratio around zero as expected (see also Supporting Information). Moreover, the polarization state of the light incident on the sample is also characterized, by rotating the linear polarization of the light coupled into the SNOM fiber using a $\lambda/2$ waveplate, and measuring the intensity transmitted by a linear polarizer used as sample for each position of the $\lambda/2$ waveplate (see also Supporting Information). We have measured a ratio between the maximum and the minimum intensity transmitted by the polarizer in the range of $10^1$-$10^2$. Overall, calibration experiments allow any contribution of the measurement set-up to the obtained results to be ruled out.

RESULTS AND DISCUSSION

Figure 1 shows SEM pictures of MEH-PPV fibers produced by electrospinning solutions with different polymer concentrations. The fibers generally feature a ribbon shape, with average width in the range 500-600 nm and width:height ratio of about 10:1. The average fiber width increases by roughly 30% upon increasing the polymer concentration in the 70-200 μM range. In addition, fibers emit bright light, allowing the chain order to be investigated by optical methods (Fig. 2a). Figure 2b shows the confocal transmission micrograph of excitation laser light, collected by crossed polarizers (analyzer axis perpendicular to the incident laser polarization), for nanofibers positioned at 0°, 65° and 90° with respect to the incident laser polarization. A significant transmitted signal can be measured only for fibers positioned at 65°, indicating optical anisotropy which is expectedly the result of a preferential molecular alignment along the fiber length.

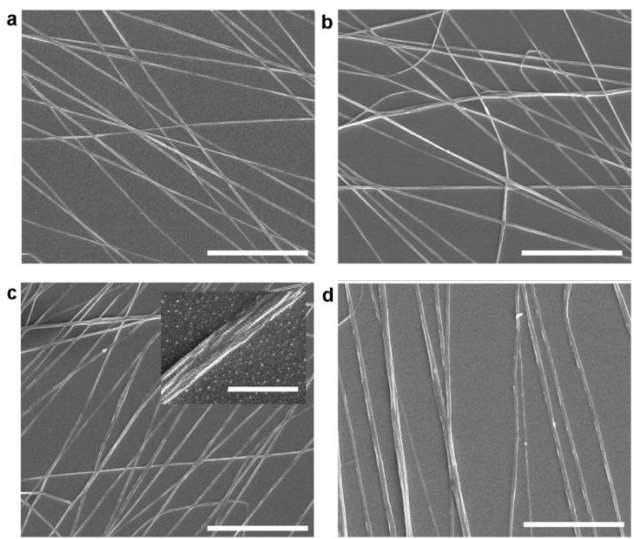

**Figure 1**. SEM images of electrospun MEH-PPV fibers realized by varying the solution polymer concentration in the range 70-200 μM. The corresponding polymer volume fraction, $\phi$,

is 0.025 (a), 0.036 (b), 0.054 (c) and 0.064 (d), respectively. Scale bar: 20 μm. Inset in (c):

Zoomed micrograph of an individual fiber highlighting its ribbon shape. Scale bar: 2 μm.

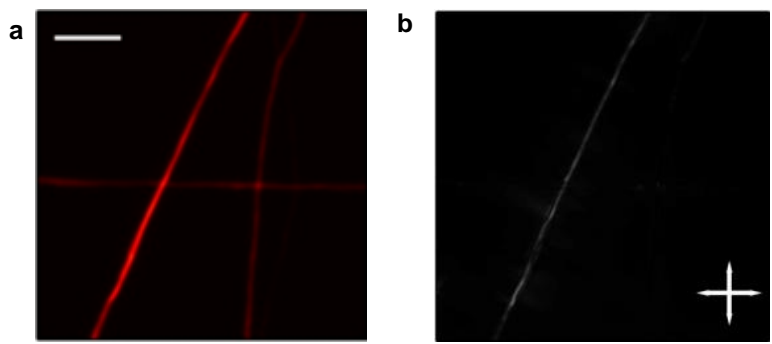

**Figure 2**. (a) Fluorescence confocal micrograph of conjugated polymer fibers. Scale bar: 10

μm. (b) Confocal map of the exciting laser intensity transmitted by the fibers, collected simultaneously to the emission map in (a). The polarization of the excitation laser (highlighted by the horizontal arrow) is aligned parallel to the longitudinal axis of the horizontal fiber, whereas the axis of the analyzer (highlighted by the vertical arrow) is positioned perpendicularly to the incident laser polarization.

Polarized near-field absorption microscopy (Fig. 3a) provides a direct measurement of the spatial variation of polymer alignment, through the map of the local dichroism, γ, i.e. the normalized difference between the transmission of radiation polarized along two mutually orthogonal directions. The map (Fig. 3b) is determined by the distribution and anisotropy of absorbing chromophores (see Supporting Information). Here, the most important finding is the spatial variation of dichroism and, consequently, of molecular alignment (Fig. 3c).

Unexpectedly, the sign of the dichroic ratio, γ, is not constant throughout the fiber, because of regions showing preferential absorption of light polarized along or across the fiber axis (for the scan shown, they correspond to negative or positive γ, respectively).

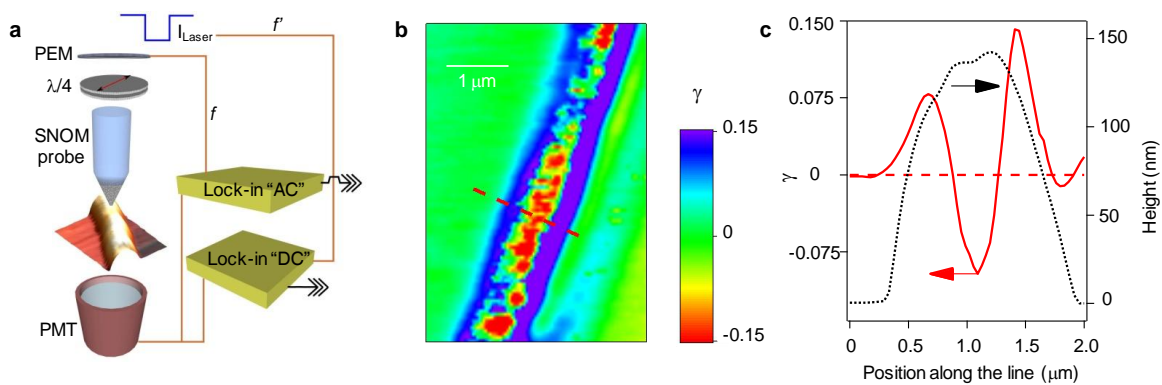

**Figure 3**. (a) Schematics of the polarization-modulation SNOM measurement. PEM: photoelastic modulator, PMT: photomultiplier. (b) Map of the dichroic ratio of a single MEH-PPV fiber. The dichroic ratio is zero for non-optically active regions (background contribution subtracted, see Supporting Information). (c) Line profile analysis displaying the cross-sections, along the dashed segment in (b), of γ (continuous line) and of topography got simultaneously with the optical data (dotted line). The change in sign of γ when crossing the fiber (dashed horizontal line corresponding to γ= 0) indicates different alignments of the polymer with respect to the fiber axis.

These results suggest the presence of a core, with width ~40% of fiber diameter, where chromophore dipoles preferentially align along the fiber length, whereas molecules closer to the fiber border show a preferential radial orientation. The decrease of the dichroic ratio from positive to null values nearby the fiber edges (i.e. for positions roughly $\leq 0.5$ μm and $\geq 1.5$ μm according to the horizontal axis of Fig. 3c) may be, instead, affected by the local curvature of the ribbon-shaped fibers and will be not considered in the following analysis. To learn more about the origin of the found spatial variation of the molecular alignment, we use a model of the polymer network and perform simulations of its dynamics, as previously developed for fully flexible and semiflexible polymer chains.[24-26] Simulations are here aimed at better rationalizing the observed chain orientation in the core, and at assessing the relevant process variables determining such orientation, thus ultimately allowing the macroscopic physical properties of the electrospun fibers to be tailored and controlled. In this approach, a flexible polymer chain is modeled as a series of $N$ rigid segments, each of length $b=nd$ ($n$ spherical beads of diameter $d \cong 1.2$ nm, each bead consisting of 2 chemical monomers). The segment length $b$ represents the average distance between two neighboring bonding defects along the chain backbone, where a bonding defect introduces local flexibility in the chain.[27] The corresponding defects concentration, using two chemical monomers per bead, is $(2n)^{-1}$ of monomers. The chain conformational correlation is lost above the scale of a segment due to the bonding defects, and therefore the rigid segment $b$ is regarded as a Kuhn segment, and a freely jointed chain model is assumed. Fully flexible polymers are a particular case of the model ($n=1$),[24] and generality is retained by using the segmental aspect ratio parameter, $n$, to specify the degree of chain flexibility.

In general, the high entanglement of chains creating a connective network determines the viscoelastic property of semidilute solutions. An entanglement can be simply defined as a topological constraint that inhibits intercrossing of two chains. The conformation of the entangled polymer network in the semidilute solution and the interactions relevant to the solvent type are described by scaling laws. When the segmental aspect ratio is high, an entanglement strand (i.e., a chain section between two adjacent entanglements) has the same length scale as the network correlation length (mesh size), $\xi$, the end-to-end distance of an unperturbed subchain containing $N_s$ rigid segments. Given the aspect ratio $n$, polymer volume fraction $\phi$, and solution properties expressed by Flory's exponent,[28] $\nu$ and Flory's interaction parameter, $\chi$, the number of rigid segments in a subchain for good solvents is (Supporting Information):

$$N_s \approx \left(\frac{n}{1-2\chi}\right)^{3(2\nu-1)/(3\nu-1)} \left(n^2\phi\right)^{-1/(3\nu-1)},\qquad(1)$$

and the corresponding correlation length is $\xi \approx b\left[(1-2\chi)/n\right]^{2\nu-1}N_s^{\nu}$.[29]

The mapping of $N_s$ as a function of $n^2\phi$, for different solvent qualities, is depicted in Fig. 4a. The effect of various solvents is discussed in detail in the Supporting Information. When the calculated $N_s$ is above the $N$ limit (upper dotted line, designating the overlap concentration $\phi^*$), the polymer network is not sufficiently entangled for elastic stretching. The limit $n$=1 (lower dotted line) designates the minimal selectable $n$ value. For our solvent mixture (THF:DMSO 4:1 wt:wt), the interaction parameter can be estimated as $\chi \cong 0.38$,[30] and for our polymer volume fraction, $\phi = 0.025$, the transition from ideal to real chain conformation occurs at $n \cong 2.7$ beads (point B' in Fig. 4a). The corresponding defects concentration (19% of monomers) is much higher than typical values (<10%, equivalent to $n$>5),[27] and therefore the conformation of subchains is close to ideal ($\theta$-solvent line in Fig. 4a, right to point B'). At the high temperature limit (athermal solvent), the transition from ideal to real conformation occurs at $n \cong \phi^{-1/3} \cong 3.4$ beads (point B in Fig. 4a), equivalent to 15% defects. Thus, as a chain is stiffer (higher $n$) it is more likely to be practically ideal, regardless of the solvent quality, provided that sufficient entanglement exists. At low concentration, when $n < \phi^{-1/2}$ (left to point A in Fig. 4a), the subchain consists of many segments and does not interact with other chains [Fig. 4b(i)]. When

$n \approx \phi^{-1/2}$ (point A in Fig. 4a), the correlation length $\xi$ is of the same scale as the segment

length $b$ [Fig. 4b(ii)].

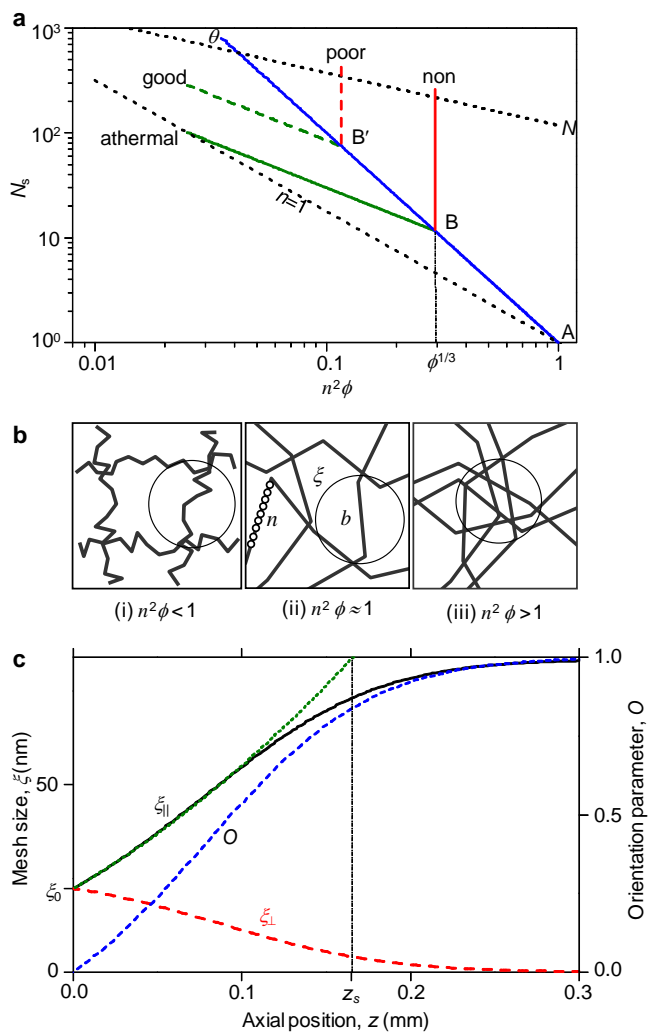

**Figure 4.** (a) Plot of $N_s$ *vs.* $n^2\phi$ and solvent quality. The $\theta$-solvent curve marks the crossover

between good and poor solvents. The dotted lines constitute the upper and lower limits for $\phi$=

0.025. Polymer molecular weight =380,000, equivalent to $N_{beads} = 730$. Points B and B', plotted for $\phi$=0.025 for Flory's interaction parameter $\chi \cong 0$ and $\chi \cong 0.38$, respectively, mark the transition from ideal subchains (right) to real subchains (left). Prefactors are omitted for sake of simplicity. (b) Crossover [Point A in (a)] of the polymer network conformation with respect to the scale of the correlation length, $\xi$ (circles) and the segment length, $b$: (i) regular semi-dilute, $\xi > b$, (ii) crossover, $\xi \approx b$, and (iii) different chains intermix within a single correlation volume, $\xi \approx b$. (c) Simulation of subchains during electrospinning. The axial mesh size $\xi_\parallel$, radial mesh size $\xi_\perp$, and orientation parameter $O$ are plotted *vs.* the axial position, $z$, along the jet. $\xi_\parallel$ is compared to the theoretical model (dotted line). The position close to full subchain extension is designated by $z_s$. Parameters used: ideal chain, $\phi = 0.025$, $n = 5$ beads, $d = 1.2$ nm, $\xi_0 \cong 20$ nm, $N_s = 14$ segments. Jet dynamics is from Fig. S4 (see Supporting Information).

However, when chains are not fully flexible, the correlation volume is not completely occupied by a single segment, and further increase of $n$ and/or $\phi$ is possible. The network then crosses over to a state where different chains intermix within a single correlation volume [Fig. 4b(iii)], increasing the probability of interchain overlap. The increased interaction between neighboring rigid segments may lead to nematic ordering and enhanced orientation, according to Onsager theory. For the volume fraction used in the experiment, $\phi$=0.025, this crossover occurs at $n \cong 8.6$ beads, corresponding to defects concentration of ~6% of monomers.

On these bases, the evolution of the polymer conformation under dynamic tension can be described by a beads-and-spring lattice model and a 3D random walk simulation.[24] To this aim, the calculated number of segments per subchain, $N_s$, the initial correlation length, $\xi_0$, and the pertaining experimental conditions are used as input to the simulations. The jet velocity is derived from the measured jet radius $a$, subjecting each subchain to a hydrodynamic force induced by the solvent, as well as to entropic elastic forces applied by its neighboring subchains.

The simulation results are presented in Fig. 4c. It is seen that subchains fully extend within less than 1 mm from the jet start (position $z_s$), while contracting laterally, and their segments become fully oriented along the jet axis. The theoretical expression for the axial stretching (dotted line in Fig. 4c), derived for linear elasticity, is given by:[24] $\xi_{II} \approx (\nu/\nu_0)\xi_0 = (a_0/a)^2 \xi_0$ , whereas the orientational parameter (dashed line in Fig. 4c) is defined as: $O = \frac{3}{2}\left\langle \cos^2\theta \right\rangle - \frac{1}{2}$, $\theta$ being the angle between a rigid segment of the polymer molecule and the longitudinal axis. An example of the conformational evolution of a single subchain under the same conditions is shown in Fig. 5.

The polymer chain is entangled with other chains in the solution (Fig. 5a-b). Each subchain (an entanglement strand) starts from an equilibrium conformation at the jet start (Fig. 5c), proceeds through intermediate stretching, and approaches full extension and lateral contraction (Fig. 5d). The subchain conformation is sensitive to the average concentration of defects. A change in the defects concentration from 10% of monomers to 15% raises the subchain size from 14 to 67 segments and increases the correlation length. The lateral contraction of individual subchains affects the conformation of the whole polymer network, narrowing its radius $a_p$ faster than the narrowing of the jet radius $a$. The dominant effect is of axial stretching and lateral contraction, resulting in compacting of the network toward the jet core. Thus, the model and simulation predicts axial alignment at the jet center, while closer to the jet boundary the stretching effect is not dominant and other mechanisms prevail. These may include the influence of surface charges as recently investigated in polyamide 6 nanofibers.[31] These charges can be tailored by changing the polarity of the applied voltage,[32] and can lead to distinct properties such as enhanced surface energy compared to solution-processed films.

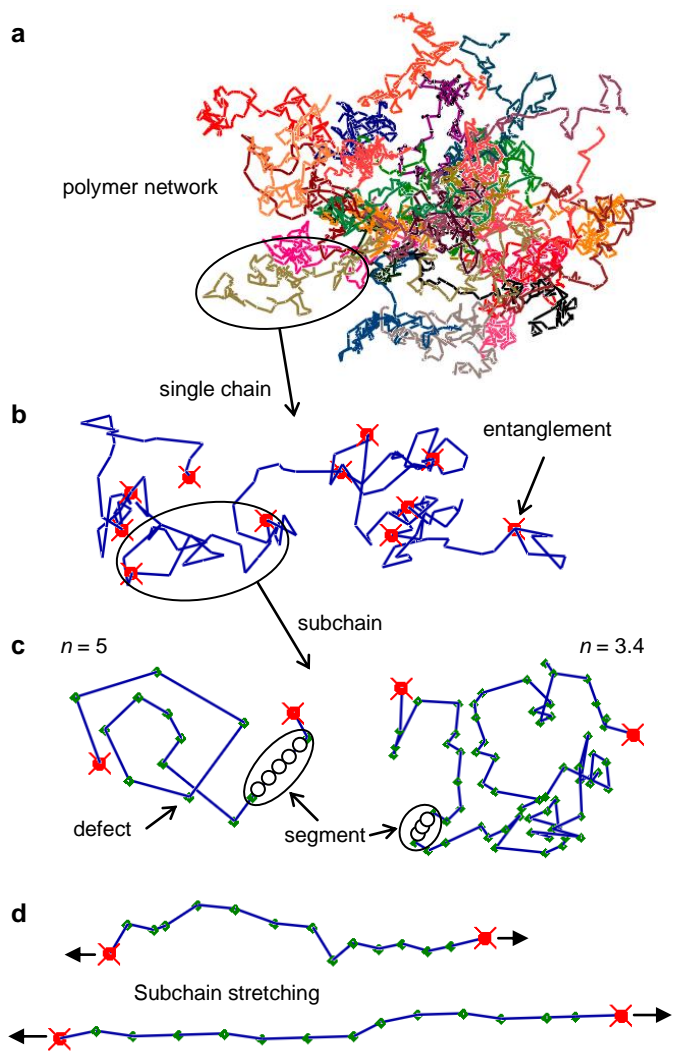

**Figure 5**. (a) Polymer network at rest. (b) A single chain with $N = 146$. (c) Examples of single subchains, left $N_s = 14$ ($n = 5$, 10% defects), right $N_s = 67$ ($n = 3.4$, 15% defects). (d) Stretched subchains, $N_s = 14$, $z = 0.08$ mm and $z = z_s = 0.16$ mm.

In our model, full extension is approached when $\xi_{\text{II}} \approx bN_s$, i.e. at the axial position $z_s$ and a corresponding jet radius $a_s$. The axial position of full stretching, omitting the effect of $n$, scales as (Supporting Information):

$$\frac{z_s}{a_0} \sim N^{3/2} \begin{cases} \phi^{\frac{\nu+2}{2(3\nu-1)}} & \text{athermal solvent} \\ \phi^{11/6} & \theta\text{-solvent.} \end{cases} \qquad (2)$$

The estimated $z_s$ for various solvent qualities, allowing the axial position of full stretching and consequently the resulting fiber properties for each particular nanofabrication experiment to be predicted, is shown in Fig. 6a. Typically $z_s$<1 mm and $a_0/a_s = 2-10$ , close to the jet start. Considering that the final radius reduction ratios in electrospinning are typically $10^2$-$10^4$, substantial stretching occurs quite early in the process. For given polymer concentration and molecular weight, when $n$ is larger (i.e. longer segments, equivalent to lower defects concentration), full stretching is approached at a higher jet radius and lower $z_s$; however, at the same time, the number of entanglements per chain $N/N_s$ is higher and therefore the solution viscosity will be larger, increasing $z_s$. In contrast to the radius reduction ratio $a_0/a_s$, the axial position $z_s$ is strongly affected by the jet rheology, resulting in a concentration dependence with a large positive exponent, as well as added dependence on the molecular weight (Fig. 6a). In addition to its dependence on the molecular weight and concentration as expressed in Eq. (2), $z_s$ strongly depends on the intensity of the electrostatic field $E$ and the jet initial velocity $v_0$. It can be shown that this dependence may be approximated by $z_s \sim v_0^{1/2} E^{-1}$, meaning that a high strain rate, caused by high $E$ and low $v_0$, should result in earlier stretching.

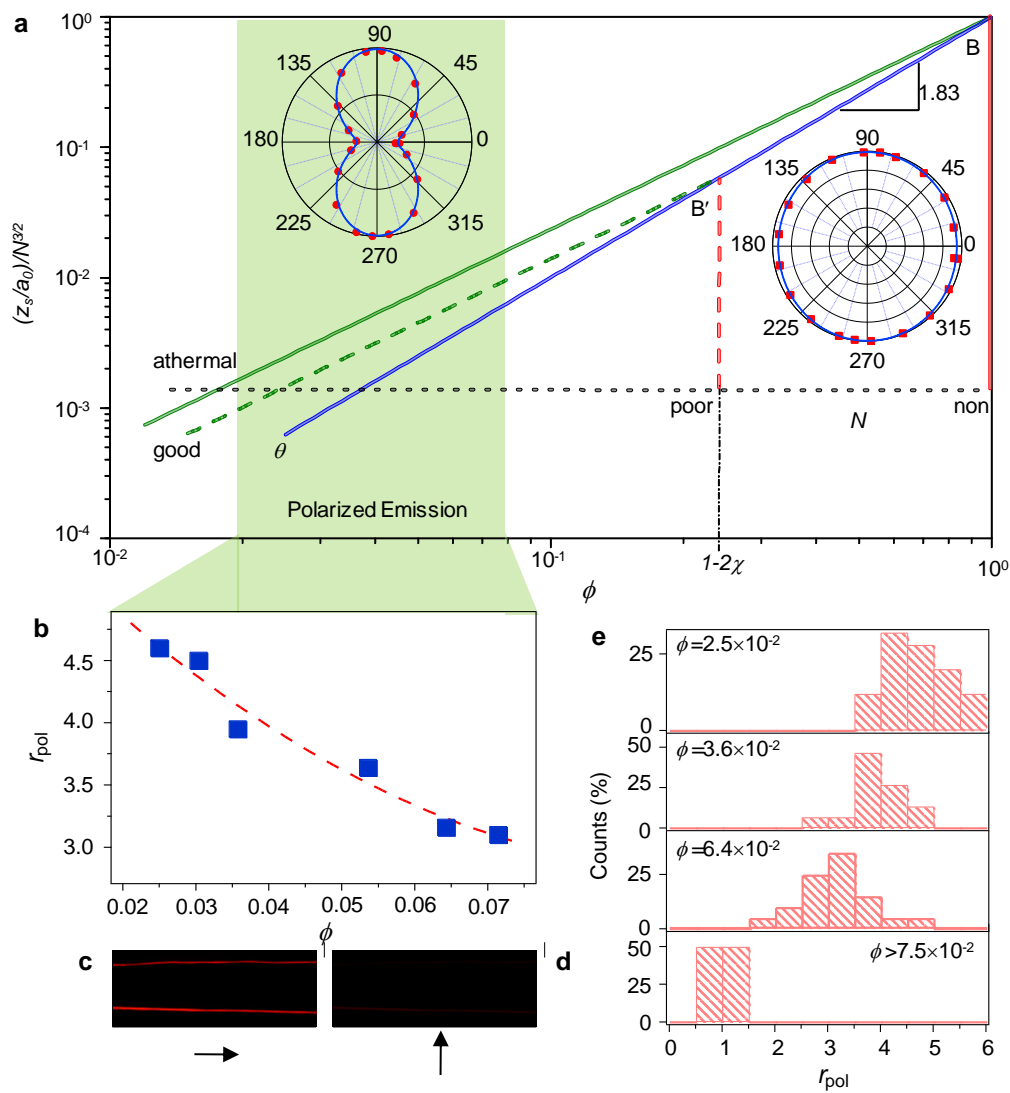

**Figure 6**. (a) Plot showing the axial position where subchains approach full extension, $z_s / a_0$, normalized by $N^{3/2}$, *vs*. the polymer volume fraction $\phi$ and solvent quality. The dotted line constitutes the lower limit imposed by $N_s < N$. Prefactors are omitted for sake of simplicity. Points B and B' are explained in Figure 4a. Insets: Plot of the normalized nanofiber emission intensity *vs*. the angle between the fiber and the analyzer axis, measured on fibers electrospun from a solution with $\phi = 0.03$ (left inset) and on a sprayed film for comparison (right inset). (b) Polarization ratio, $r_{pol}$ *vs*. solution volume fraction, $\phi$. The dashed line is a guide for the eyes. An unpolarized sample (sprayed film) has $r_{pol}$ = 1. (c-d) Confocal images of nanofiber polarized emission. The laser-excited emission is filtered through an analyzer with axis (highlighted by arrows) parallel and perpendicular to the fiber axis, respectively. (e) Experimental distributions of the nanofiber polarization ratio, $r_{pol}$, at different polymer concentrations.

A substantial axial stretching of chains is therefore predicted during the initial stage of the elongational flow, causing lateral contraction of the polymer network toward the center of the jet, as well as orientation of chain segments along the jet axis. Similar results obtained for fully flexible chains (particular case with $n$=1) have been confirmed by X-ray imaging of high strain rate electrified jets.[24,25] In electrospinning, the electric field provides the flow of the semi-dilute solution with a characteristic increasing velocity along the jet axis, with a strain rate that continuously increases the elastic stretching of the polymer network and reduces network relaxation. Our model shows that full chain stretching is approached at a higher jet radius as the chain is stiffer, at a region where the mass loss rate due to evaporation is still low.

When stretching is less dominant (e.g. at low electric field and high flow rate), the rapid solvent evaporation can adversely affect the polymer matrix, creating a porous nanofiber structure. Dominant evaporation can also lead to a rapid solidification of the jet surface, limiting further solvent loss from the core.[33-35] The presence of residual solvent content in the jet core would allow for chain relaxation, thus disfavoring the retention of alignment in the core as recently reported for electrospun polyvinyl-alcohol fibers.[36] Instead, when stretching is dominant as in the present case, the polymer network compacts towards the center, producing an increase of the density close to the jet axis.[26] The here observed ribbons (Fig. 1) are hence likely affected by concurring effects rather than jet skin collapse, such as flattening and relaxation processes occurring at the impact onto the substrate. This would be consistent with the presence of a slowly evaporating solvent component, i.e. with a jet time of flight which is comparable with the drying timescale,[35, 37] and supported by the joints observed in SEM micrographs of intersecting deposited fibers (e.g. Fig. 1b,c).

This description is in agreement with polarization modulation measurements (Fig. 3), showing a change in the sign of the dichroic ratio along the fiber radius, and indicating a preferred axial alignment of molecules at the fiber core, whereas molecules closer to the fiber boundary possess a preferred radial alignment. Thus, at the jet center axial stretching is dominant, and one can anticipate a propensity for interchain interaction and $\pi$-$\pi$ stacking and consequently high extent of local crystallinity.[38] At the boundary region of the jet, where the polymer concentration is reduced,[25, 26] entanglement may be low or nonexistent, allowing partial or full relaxation of chains back to their coil-shaped equilibrium state. This mechanism is also supported by a recent study on entanglement loss in extensional flow,[39] showing that the electrospinning process causes partial untangling of the polymer network when stretching is faster than the chains relaxation time.

Overall, full extension of the network occurs at an earlier stage of the jet (lower $z_s$) if the solution initial concentration, the polymer molecular weight, and the solvent quality are lower, accounting for lower network entanglement. Under such conditions, the likelihood that the extended conformation, and the associated axial molecular alignment, will partially remain in the polymer structure after solidification is higher. This enables tailoring the physical properties of fiber, such as the far-field, macroscale emission from fibers. The $z_s(\phi)$ diagram of Fig. 6a clearly relates the chain alignment, and hence the resulting polarization, to the polymer volume fraction (i.e. solution concentration). Indeed, by our approach we obtain a fine tuning of the polarization ratio of the fiber emission ($r_{pol} = I_\parallel / I_\perp$, given by the ratio between the photoluminescence intensity parallel, $I_\parallel$, and perpendicular, $I_\perp$, to the fiber axis, respectively), increasing up to about 5 by gradually decreasing the solution concentration down to a volume fraction $\phi$=0.025, as shown in Fig. 6b-e. This demonstrates the possibility of tailoring specific fiber properties by the relevant process parameters. The ultimate $r_{pol}$ values may also benefit from the higher density in the core,[26] as well as from electronic energy-transfer mechanisms, which strongly affect the emission properties of conjugated polymers.[40,41] Conjugated polymer fibers frequently show red-shifted absorption compared to spincast films (Fig. S1a and Ref. 42), a property consistent with a longer effective conjugation length consequence of the stretched conformation. In fact, the elongational dynamics of solutions leads to extended structures having interchain alignment. The bonding defects concentration, $(2n)^{-1}$ of monomers, determines chain flexibility, and appears as a possible key factor in controlling the desired morphology.

CONCLUSIONS

In summary, anisotropy at nanoscale is investigated in polymer fibers by polarization modulation SNOM absorption measurements, evidencing a variation of molecular orientation from axial to radial upon moving from the fiber axis to its surface. Modeling the evolution of the conformation of the chains network allows us to identify key parameters for controlling molecular alignment, as demonstrated by the fine control of the emission polarization. The found complex internal structure and assessment of the key influencing process variables open new perspectives for tailoring the molecular morphology and resulting fiber properties.

AUTHOR INFORMATION

**Corresponding Author**

*E-Mail: Andrea Camposeo: andrea.camposeo@nano.cnr.it,

Israel Greenfeld: green_is@netvision.net.il,

Dario Pisignano: dario.pisignano@unisalento.it

**Author Contributions**

‡ These authors contributed equally.

ACKNOWLEDGMENT

We gratefully acknowledge the financial support of the United States-Israel Binational Science Foundation (BSF Grant 2006061), the RBNI-Russell Berrie Nanotechnology Institute, and the Israel Science Foundation (ISF Grant 770/11). The research leading to these results has received funding from the European Research Council under the European Union's Seventh Framework Programme (FP/2007-2013)/ERC Grant Agreement n. 306357 (ERC Starting Grant "NANO-JETS"). The authors also gratefully thank S. Pagliara for sample preparation, E. Caldi for assistance in the SNOM measurements, S. Girardo for imaging of the polymer jet and V. Fasano for confocal images.

# SUPPORTING INFORMATION

# The conformational evolution of elongated polymer solutions tailors the polarization of light-emission from organic nanofibers

*Andrea Camposeo[1][‡][*], Israel Greenfeld[2][‡][*], Francesco Tantussi[3,4], Maria Moffa[1], Francesco Fuso[3,4], Maria Allegrini[3,4], Eyal Zussman[2], Dario Pisignano[1,5,*]*

[1]National Nanotechnology Laboratory of Istituto Nanoscienze-CNR, via Arnesano, I-73100 Lecce (Italy)

[2]Department of Mechanical Engineering, Technion – Israel Institute of Technology, Haifa 32000 (Israel)

[3]Dipartimento di Fisica "Enrico Fermi" and CNISM, Università di Pisa, Largo Bruno Pontecorvo 3, I-56127 Pisa (Italy)

[4]Istituto Nazionale di Ottica INO-CNR, Sezione di Pisa, I-56127 Pisa (Italy)

[5]Dipartimento di Matematica e Fisica "Ennio De Giorgi", Università del Salento, via Arnesano I-73100 Lecce (Italy)

*Corresponding Authors: E-mail: andrea.camposeo@nano.cnr.it, green_is@netvision.net.il, dario.pisignano@unisalento.it

‡These authors contributed equally

## 1. Theoretical modelling rational

The distinctive photophysical properties of conjugated polymers are strongly affected by the structural conformation of the polymer solid matrix,[S1-S5] more specifically by the ordering and orientation of chain sections, typically with a conjugation length of ~10-15 monomers in MEH-PPV.[S1,S6] Individual conjugated polymer chains in a dilute solution can assume various conformations, and hence different optical properties, depending on solvent quality and bonding defects. Theoretically, when defects are not present, the chain is semi-flexible and takes the form of a toroid or rod. However, chemical defects introduced by polymer synthesis and reactivity substitute conjugated links by tetrahedral links along the chain backbone, creating defect coil or cylindrical shapes as a result of the increased flexibility. The defects concentration is in the order of 2.6% to 10% of monomers.[S1,S2,S5] Generally, in good solvents conjugated chains are swollen and polymer-solvent interactions are dominant, whereas in poor solvents polymer-polymer intrachain and interchain interactions are stronger, favoring aggregation and π-π stacking, respectively.[S2-S4]

The processing of thin conjugated polymer films from dilute solutions by spin coating, dip coating and casting has a relatively low impact on chain conformations in view of the weak dynamics. Such conformations, extensively investigated both theoretically and experimentally,[S1-S5] are partly retained after solvent evaporation due to memory effect[S4] but are still close to the chain equilibrium state. By contrast, electrospinning of semidilute solutions of conjugated polymers is characterized by strong stretching generated by the high electrostatic field, resulting in extended chain conformations and longer conjugation lengths, evidenced by red-shifted optical absorption of fibers and smaller phase separation length scale.[S7] It is therefore a specific aim of the present study to provide a modeling approach for the conformational evolution of the conjugated polymer chains during electrospinning, including its dependence on solution properties and jet dynamics. It is further intended to describe the impact of the modeled conformation on the solid nanofiber microstructure and photophysical properties in light of MEH-PPV optical observations.

The model is generalized by tuning the degree of chain flexibility with the segmental aspect ratio parameter, and therefore applies to a wide range of linear flexible polymers, including to conjugated polymers with different levels of defects concentration, as well as to fully flexible polymers which are a particular case of the model.

## 2. Chain conformation in solution

Considering a polymer chain having $N$ rigid segments, each of length $b = nd$ ($n$ spherical beads of diameter $d$). In a semi-dilute solution, a chain section (subchain) within a correlation volume is essentially unperturbed by other chains, and therefore its end-to-end distance for good solvent can be expressed by Flory's radius:

$$\xi \approx b\left(\frac{\text{v}}{b^3}\right)^{2\nu-1} N_s^{\nu}. \tag{S1}$$

where $\text{v}$ is the excluded volume and $\nu$ is Flory's exponent. Remembering that a correlation volume encloses a single subchain and that correlation volumes are space filling, the polymer volume fraction in the solution is:

$$\phi \approx \frac{N_s n d^3}{\left(2\xi/\sqrt{6}\right)^3} = \left(\tfrac{3}{2}\right)^{3/2} N_s n^{-2}\left(\frac{b}{\xi}\right)^3. \tag{S2}$$

The correlation length is obtained by substituting $N_s$ from Equation (S2) into Equation (S1):

$$\xi \approx b \left( \frac{b^3}{v} \right)^{(2\nu-1)/(3\nu-1)} \left( n^2 \phi \right)^{-\nu/(3\nu-1)}, \tag{S3}$$

with a pre-factor of order unity, $\left( \frac{3}{2} \right)^{3\nu/(6\nu-2)}$. This relation can be obtained also by de Gennes scaling approach. The corresponding number of segments is:

$$N_s \approx \left( \frac{b^3}{v} \right)^{3(2\nu-1)/(3\nu-1)} \left( n^2 \phi \right)^{-1/(3\nu-1)}. \tag{S4}$$

with a pre-factor of order unity, $\left( \frac{3}{2} \right)^{3/(6\nu-2)}$. The excluded volume of a non-spherical segment can be expressed by $v \approx b^2 d (1 - 2\chi)$, where $\chi$ is Flory's interaction parameter, and therefore:

$$\frac{b^3}{v} \approx \frac{b^3}{b^2 d (1 - 2\chi)} \approx \frac{n}{(1 - 2\chi)}. \tag{S5}$$

The final expression for the number of segments for good solvent is:

$$N_s \approx \left( \frac{3}{2} \right)^{3/(6\nu-2)} \left( \frac{n}{1-2\chi} \right)^{3(2\nu-1)/(3\nu-1)} \left( n^2 \phi \right)^{-1/(3\nu-1)}, \tag{S6}$$

which reduces to $N_s \approx \left( \frac{3}{2} \right)^3 \left( n^2 \phi \right)^{-2}$ segments for an ideal chain ($\nu = 1/2$).

Figure 4a displays the dependence of $N_s$ on $n^2 \phi$, for different solvent types. The $\theta$-solvent curve ($\nu = 1/2$, $\chi = 1/2$) marks the crossover between good and poor solvents, for which the subchain is a random walk of segments (ideal chain conformation). In good solvents ($\nu = 3/5$, $\chi < 1/2$), the subchain is swollen (real chain conformation) and therefore the values of $N_s$ and $\xi$ are lower. The corresponding curve starts at point B', and the athermal limit curve ($\chi = 0$) starts at point B. In poor solvents ($\nu = 1/3$, $\chi > 1/2$), the subchain is shrunk and therefore $N_s$ and $\xi$ are higher, up to the non-solvent limit ($\chi = 1$) where subchains are fully collapsed and phase-separated from the solvent. The corresponding curves start at point B' and B, respectively. $N_s$ is smaller at high aspect ratio (low defects concentration) and is proportional to $n^{-(5-6\nu)/(3\nu-1)}$, i.e. $n^{-4}$ for a $\theta$-solvent and $n^{-1.75}$ for good and athermal solvents.

For chain sections where the excluded volume interaction energy is weak, the conformation is dominated by the thermal energy and is therefore diffusive and ideal. The transition from real to ideal conformation occurs at the length scale of a thermal blob possessing $b^6 / v^2$ segments. Thus, the number of segments in a thermal blob is:

$$N_{sT} \approx \frac{b^6}{v^2} \approx \left[ \frac{n}{1-2\chi} \right]^2 , \qquad (S7)$$

which reduces to $N_{sT} \approx n^2 \approx \phi^{-2/3}$ for athermal solvents ($\chi = 0$). For solvents other than a $\theta$-solvent, point B' in Figure 4a marks the position where the subchain is of the same length scale as the thermal blob. For larger $n$ (smaller $N_s$), excluded volume interactions are weak and the thermal energy is dominant; the $\theta$-solvent curve applies and subchains have an ideal conformation. For smaller $n$ (larger $N_s$), excluded volume interactions are dominant outside the thermal blob; the good solvent or poor solvent curves apply and subchains have an intermediate swollen or collapsed conformation, respectively. The minimal thermal blob size is at the athermal limit (point B), where the athermal solvent or non-solvent curves apply and subchains have a fully swollen or collapsed conformation, respectively. Note that point B slides to the right as the volume fraction increases.

Given the molecular weight or the number of beads in the complete chain, $N_{beads}$, the chain's number of segments is $N = N_{beads}/n$. The average number of entanglements per chain, $N/N_s \sim n^{3(2-3\nu)/(3\nu-1)}$, decreases for smaller $n$ (i.e. higher defects concentration and chain flexibility) and the solution may turn dilute. Thus, stiffer chains can establish a level of entanglement sufficient for electrospinning at lower concentration and molecular weight. For poor solvents, $N_s$ is close to $N$, not assuring sufficient entanglements for electrospinning.

The model uses the correlation length as representative for the entanglement length, a chain section between two consecutive entanglements along the chain. In a semi-dilute solution, the elastic modulus of the polymer network is proportional to the number density of entanglement strands. In an athermal solvent, an entanglement strand containing $N_e$ segments is related to the correlation length by $N_e \approx N_e(1)N_s$, where $N_e(1)$ is the number of segments in an entanglement strand in a melt. Although the entanglement strand length is always longer than the correlation length, in conjugated polymers they are of similar length scale because of the high segmental aspect ratio. Using the known expression for the number of chains within the confinement volume of a single strand in a melt, $P_e \approx (b^3/v_0)\sqrt{N_e(1)}$, where $v_0 \approx d^2 b$ is the segment volume, the number of strands in a conjugated polymer strand in a melt is:

$$N_e(1) \approx \left(\frac{P_e v_0}{b^3}\right)^2 \approx \left(\frac{P_e}{n^2}\right)^2. \qquad (S8)$$

Since $P_e \cong 20$ for all flexible polymers regardless of their segmental aspect ratio, $N_e(1) \approx 1$ when $n \geq \sqrt{P_e} \cong 5$, and therefore practically, for typical defects concentration, $N_e \approx N_s$ as used in the modeling.

### 3. Network stretching during electrospinning

Using Eq. (1) the expression for chain extension, $\xi_{II} \approx (v/v_0)\xi_0 = (a_0/a)^2 \xi_0$,[S8] the jet radius reduction ratio at full stretching is expressed for good solvents by:

$$\frac{a_0}{a_s} \approx \left(\frac{bN_s}{\xi_0}\right)^{1/2} \approx \left(\frac{n}{1-2\chi}\right)^{(1-2\nu)/2} N_s^{(1-\nu)/2}$$

$$\approx \left(\frac{n}{1-2\chi}\right)^{(2\nu-1)/(3\nu-1)} \left(n^2\phi\right)^{-(1-\nu)/[2(3\nu-1)]}. \tag{S9}$$

Thus, $a_0/a_s$ is reduced as the chain is stiffer, by a factor of $n^{-(2-3\nu)/(3\nu-1)}$, i.e. $n^{-1}$ for a $\theta$-solvent (for which $\nu$=1/2) and $n^{-0.25}$ for good and athermal solvents ($\nu$=3/5).

Finally, in order to find the axial position $z_s$ where subchains approach full extension, it is necessary to estimate the dependence of the velocity parameter $z_0$ in Eq. (S20) on the parameters that determine the jet rheology, predominantly the solution viscosity $\eta$. Additional influencing parameters, such as the electric field and conductivity, flow rate, and needle diameter, are assumed constant in the current analysis. Sufficiently far from the orifice ($z/z_0 \gg 1$), the jet velocity gradient is proportional to $z_0^{-2}$ (using the exponent $\beta$=1 in Eq. S20). Since, for a given tensile stress (e.g. due to electric field), the gradient is approximately proportional to $\eta^{-1}$, the velocity parameter scales as $z_0 \sim \eta^{1/2}$. Written in dimensionless parameters, $z_0/a_0 \sim \eta_{sp}^{1/2}$, where $\eta_{sp} \cong \eta/\eta_s$ is the specific viscosity ($\eta_s$ is the solvent viscosity, $\ll \eta$). Hence, The dependence of the axial position on the jet radius (see Eq. S20 in the following Sec. 6) can be approximated by:

$$\frac{z}{a_0} \cong \frac{z_0}{a_0}\left(\frac{a}{a_0}\right)^{-1} \sim \eta_{sp}^{1/2}\frac{a_0}{a}, \tag{S10}$$

where the specific viscosity can be obtained from the known expressions for semi-dilute solutions,[S9] $\eta_{sp} \approx N^3\phi^{3/(3\nu-1)}$ for an athermal solvent and $\eta_{sp} \approx N^3\phi^{14/3}$ for a $\theta$-solvent (prefactors are omitted).

Using the radius reduction ratio from Equation (S9) and the viscosity expressions, $z_s$ scales as:

$$\frac{z_s}{a_0} \sim N^{3/2}\begin{cases}\phi^{\frac{\nu+2}{2(3\nu-1)}} & \text{athermal solvent}\\[2mm]\phi^{11/6} & \theta\text{ solvent},\end{cases} \tag{S11}$$

depicted in Figure 6a. The concentration dependence of $z_s$ is $\phi^{1.63}$ for an athermal solvent and $\phi^{1.83}$ for a $\theta$-solvent. Note that this expression is written for fully flexible chains ($n=1$), with the intention to demonstrate the trends of the dependence on the concentration, solvent quality, and molecular weight.

## 4. Nanofiber optical properties

The absorption spectra are measured by using a UV/visible spectrophotometer (Lambda950, PerkinElmer) and an integrating sphere (Labsphere). PL spectra are acquired by exciting the nanofibers with a diode laser ($\lambda$=405 nm, polarization parallel to the fiber axis) and collecting the emission through a fiber-coupled monocromator (iHR320, JobinYvon) equipped with a charged coupled device (Jobin Yvon). In Figure S1a we compare the absorption spectrum of a mat of MEH-PPV nanofibers with a reference spincast film. The film displays a maximum absorption at about 510 nm, whereas a peak red shift of about 5 nm is measured for fibers, which indicates a slight increase of the effective conjugation length attributable to a more ordered molecular packing.[S7] Moreover, the absorption spectrum of the fibers is also broader compared to the film sample. The full width at half maximum (FWHM) of the fiber absorption spectrum is 150 nm, whereas a FWHM of 120 nm is measured for the film. Such broadening has been also observed in other electrospun MEH-PPV nanofibers and attributed to a more inhomogeneous environment.[S7]

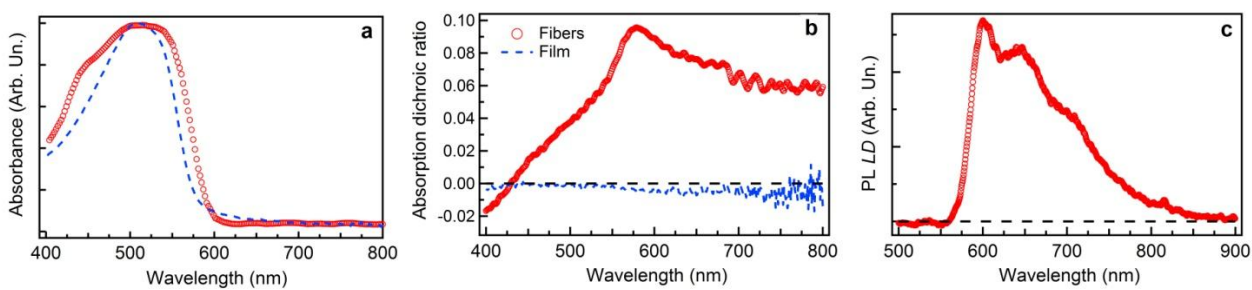

**Figure S1**. **a**, Absorption spectra of a reference MEH-PPV film (dashed line) and of the nanofibers (circles). **b**, Dichroic ratio spectra of fibers (empty circles) and reference spincast films (dotted line). The spectra are obtained by measuring the polarized absorption spectra, with incident light polarization aligned parallel ($A_{\parallel}$) and perpendicular ($A_{\perp}$) to the fiber alignment axis, respectively. The dichroic ratio is then calculated by: $\gamma = (A_{\parallel} - A_{\perp})/(A_{\parallel} + A_{\perp})$. The spectrum of the reference film clearly displays average null values. **c**, PL linear dichroism spectrum ($LD_{PL}$), calculated by PL spectra exciting the fibers with the laser polarized along the fiber axis and collecting the PL through a polarization filter with axis parallel ($I_{PL//}$) and perpendicular ($I_{PL\perp}$) to the fiber axis, respectively (i.e., $LD_{PL} = I_{PL//} - I_{PL\perp}$). The horizontal black dashed lines mark null values.

The absorption dichroic ratio spectrum is shown in Figure S1b. The plot, obtained from polarized absorption measurements performed on uniaxially aligned MEH-PPV nanofibers, evidences a predominant absorption for light polarized along the fiber longitudinal axis, a fingerprint of a preferential alignment of the polymer backbones along the fiber length. The values are peaked at about 575 nm, corresponding to the (0-0) vibronic replica of the π-π* electronic transition. Figure S1c displays the photoluminescence (PL) linear dichroism defined as the difference between the intensity of the light emitted by fibers with polarization parallel or perpendicular to the fiber axis ($LD_{PL} = I_{PL//} - I_{PL\perp}$). The measured polarization anisotropy of the emission further supports the anisotropic packing of MEH-PPV molecules in the electrospun fibers.

### 5. Near-field Optical Microscopy set-up

The analysis of the optical activity at the local scale is performed by using a polarization-modulated scanning near-field optical microscope (SNOM) developed on the basis of a multi-purpose head.[S10] The instrument operates in the emission-mode: the sample interacts with the near-field produced by a tapered optical fiber probe (50 nm nominal diameter apical aperture). Radiation is collected below the sample by an aspheric lens and directed onto a miniaturized photomultiplier. The configuration is hence similar to that of conventional optical transmission measurements, but for the use of the near-field, the key component enabling spatial resolution below the diffraction limit.

Producing maps of the linear dichroism requires the ability to measure the response of the sample upon excitation with controlled polarization states. To this aim, polarization modulation (PM) methods are typically employed.[S11] The main motivation is the long duration of SNOM scans, that makes mechanical drifts likely to occur. As a consequence, subsequent scans of the same region recorded with rotated polarizations can be hardly carried out. PM circumvents the problem by continuously manipulating the polarization state at every point of a single scan. In addition, signal-to-noise ratio is improved thanks to demodulation through lock-in amplifiers, that makes the measurement of small dichroism variations feasible.[S12]

Core of the polarization modulation system is a photoelastic modulator (PEM, Hinds Instruments PEM-100) which acts as a waveplate whose retardation is periodically modulated at $f = 50$ kHz.[S13] As shown in Fig. S2, the modulator is followed by a $\lambda/4$ waveplate. Being the linear polarization of the excitation laser directed at 45° with respect to the PEM optical axes, the polarization entering the $\lambda/4$ waveplate is periodically modified through linear to elliptical and circular states. After passage through the waveplate, oriented at -45° with respect to the PEM axes, the polarization is converted back to linear, but its in-plane direction gets now periodically modulated. In typical operating conditions, the freely adjustable, maximum retardation produced by the PEM is set to $A = \pi$. Consequently, the whole range of directions (0-360°) is spanned two times in a single modulation period $T = 1/f$. Lock-in demodulation at twice the frequency $f$ brings information (hereafter called AC signal) on the response of the sample to polarized radiation. Since the intensity of the signal collected by the photodetector can be affected by a number of effects, including instrumental drifts, variations in the coupling efficiency with the near-field, as well as polarization-independent absorption, the AC signal must be normalized. A reference can be easily obtained by averaging the photodetector output over all polarization states. In our setup this is achieved by using an independent lock-in amplifier connected at the photo-detector output and referenced to a slow ($f' < f/10$) amplitude modulation of the excitation laser, that produces a signal (hereafter called DC) time-averaged over all polarization states. The normalized AC/DC signal contains information on the linear dichroism of the sample.

Measurements at the local scale involve the use of near-field radiation. Different previous reports have demonstrated the ability of polarization modulated SNOM in achieving a polarization-related contrast mechanism and in determining the optical activity of a variety of samples.[S14-S21] However, especially when near-field probes based on tapered optical fibers are used, as in our case, special care must be devoted to account for the residual optical activity. In fact, birefringence of the fiber, as well as any other spurious effect stemming from the optical bench components (mirrors, beam splitters), can heavily affect the results, in particular their quantitative evaluation. In order to remove those unwanted contributions, prior to the measurements we analyze bare substrates, expected to show negligible optical activity. Both demodulated signals (lock-in outputs) and time-resolved data are compared to results obtained through calculations aimed at simulating the polarization-dependent behavior of the optical components in the experimental chain. Such calculations have been based on the Jones matrix formalism (more complicated methods, such as those based on the Mueller formalism, did not lead to significant advantages).

Within this frame, the time-dependent polarization of the light at the PEM output can be represented by the following vector, whose components are the electric field amplitude along two mutually orthogonal in-plane directions ($x$ and $y$) in the reference frame of the optical bench:

$$V = \frac{1}{\sqrt{2}} \begin{pmatrix} e^{i\Delta(t)} \\ 1 \end{pmatrix},$$
(S12)

with $\Delta(t) = A\sin(2\pi f t + \delta)$, where $\delta$ represents a constant phase factor accounting for instrumental delays. The modifications to the polarization produced by each component are described by $2\times 2$ matrices. The whole system behavior is then given by the product of all considered matrices.

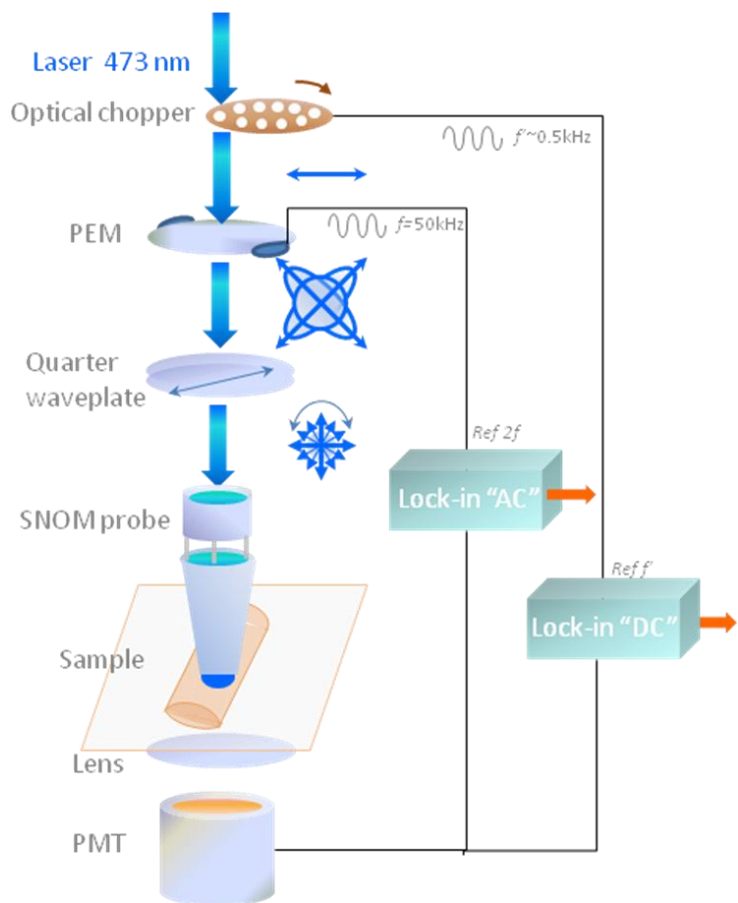

**Figure S2**. Representation of polarization modulated SNOM. Close to some of the optical components, the polarization of the excitation radiation is schematically shown. Note that, in the actual setup the optical chopper, depicted as a rotary wheel, is replaced by an acousto-optic shutter in order to prevent mechanical noise on the microscope table.

For instance, the $\lambda/4$ waveplate, aligned as in our setup, can be described by the following matrix:

$$W = \frac{1}{\sqrt{2}} \begin{pmatrix} \cos\theta & -\sin\theta \\ \sin\theta & \cos\theta \end{pmatrix} \begin{pmatrix} 1 & 0 \\ 0 & i \end{pmatrix} \begin{pmatrix} \cos\theta & \sin\theta \\ -\sin\theta & \cos\theta \end{pmatrix},$$ (S13)

with $\theta = -\pi/4$. A generic birefringent component, with its optical axes rotated by a generic angle $\beta$ with respect to the reference system, can be represented by the matrix $M$:

$$M = \frac{1}{\sqrt{2}} \begin{pmatrix} \cos\beta & -\sin\beta \\ \sin\beta & \cos\beta \end{pmatrix} \begin{pmatrix} e^{-i\Delta\phi} & 0 \\ 0 & 1 \end{pmatrix} \begin{pmatrix} \cos\beta & \sin\beta \\ -\sin\beta & \cos\beta \end{pmatrix},$$ (S14)

where $\Delta\phi$ is the optical retardation produced by the birefringent component. The matrix $M$ can conveniently describe the spurious birefringence induced by the probe.

The components of the electric field, $D$, collected by the photodetector (in the absence of any optically active sample) are given by the product:

$D = MWV,$ (S15)

which is time-dependent due to the explicit dependence of $V$ on $t$. Assuming photodetector response independent of polarization, as confirmed by specific calibration measurements showing negligible variations of the output signal for radiation polarized along two mutually orthogonal directions, the intensity $I(t) = |D(t)|^2$ can then be simulated and compared to the signal measured by the photodetector in the experiment, duly amplified and acquired by a fast digital oscilloscope averaging over many modulation cycles in order to enhance the signal-to-noise ratio. By adjusting the calculation parameters to achieve the best agreement between the simulated and the observed temporal shape of the signal, the retardation $\Delta\phi$ can be estimated.

In our experiments we select probes showing a residual birefringence, $\Delta\phi < 0.1$ rad. Moreover, since the birefringence can be altered by naturally occurring mechanical stresses of the fiber, particular care is taken to keep stable the eventual mechanical stress and to minimize long term drifts, as experimentally confirmed by measuring the AC signal fluctuations in scans of bare substrates (measured well below 10% on the duration of a whole scan, typically lasting for several tens of minutes).

In order to have an additional confirmation of the suitability of the selected probes, further tests are carried out in which a linear polarizer is used as SNOM sample. Then, we replace the PEM and $\lambda/4$ assembly with a $\lambda/2$ waveplate. Rotating such waveplate, we can obtain a linear polarization aligned along a variable direction at the entrance of the probe, and use the photodetector output to carry out an analysis mimicking the crossed-polarizer configuration. The probes used in our experiments feature extinction ratios of the order of $10^1$-$10^2$, ruling out possible polarization scrambling effects.

Once the spurious birefringence has been characterized, the dichroism of the sample can be determined by using a matrix $S$ describing a generic linearly dichroic material with its axes rotated at a generic angle $\alpha$:

$$S = \begin{pmatrix} \cos\alpha & -\sin\alpha \\ \sin\alpha & \cos\alpha \end{pmatrix} \begin{pmatrix} e^{-k_1 z} & 0 \\ 0 & e^{-k_2 z} \end{pmatrix} \begin{pmatrix} \cos\alpha & \sin\alpha \\ -\sin\alpha & \cos\alpha \end{pmatrix},$$ (S16)

where $k_1$ and $k_2$ are the absorption coefficients for radiation polarized along two mutually orthogonal directions representing the optical axes of the sample, and $z$ is the thickness of the absorbing material, that can be experimentally inferred by the topography maps. The response of the whole system produces the time-dependent electromagnetic field $D_S$:

$$D_S = SMWV.$$ (S17)

In order to compare with experimental data (at a fixed point of the scan, to remove the dependence on $z$), lock-in demodulation at frequency $2f$ (or $f$) of the intensity $I_S(t) = |D_S(t)|^2$ measured by photodetector can be easily simulated, corresponding in practice to the measured AC/DC ratio. Thanks to the use in the experiment of dual lock-in amplifiers, any possible dephasing between the reference and the signal modulation, for instance caused by the parameter $\delta$ (see Eq. S12), can be neglected.

The absorption coefficients can in turn be linked to the dichroic ratio of the material:

$$\gamma = \frac{I_{||} - I_{\perp}}{I_{||} + I_{\perp}} \tag{S18}$$

where

$$I_{||} = I_0 e^{-k_1 z} \text{ and } I_{\perp} = I_0 e^{-k_2 z} \tag{S19}$$

We note that the above defined dichroic ratio can get either positive or negative values, depending on whether $k_1 < k_2$, or vice versa. The ability to retrieve the sign of $\gamma$ is indeed very useful when the spatial distribution of the dichroic behavior in isolated systems has to be investigated, as in our case. In fact, this allows to identify regions with radically different polarization-dependent absorption, as due, for instance, to peculiar alignment of molecules.

The procedure leads to a calibration curve where the simulated AC/DC ratio is plotted as a function of $\gamma$ that can be used to put a scale bar into the experimentally produced maps. Figure S3 shows an example of the resulting curve. We estimate the uncertainty of our procedure, accounting for all the error sources in our simple calculation model, of the order of ±10%.

As shown in Fig. S3, non-zero AC/DC is expected at $\gamma = 0$. The occurrence of such a background pedestal, which is an obvious consequence of the spurious optical activity of the optical bench, is well confirmed in the experimental maps, where the bare substrate actually displays non-zero AC/DC. Thanks to the calibration curve, the effect disappears in the maps calibrated in units of $\gamma$, as shown in Fig. 3d.

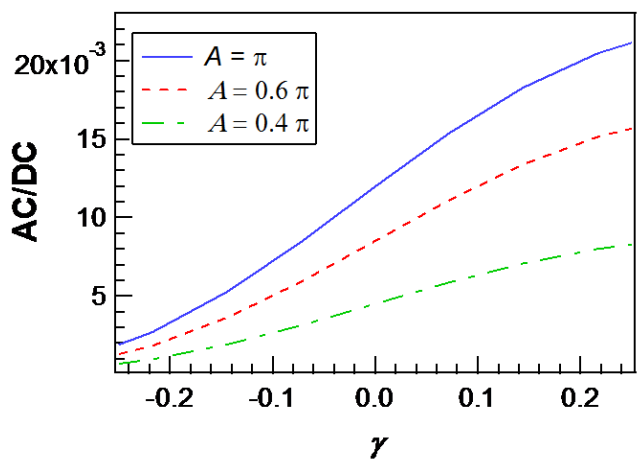

**Figure S3**. Example of a calibration curve, as defined in the text, relating the simulated AC/DC ratio with the dichroic ratio of the material. In this example, which fits experimental data, the residual birefringence of the probe is computed as $\Delta\phi = 0.06$ rad and the angle $\alpha$ appearing in Eq. S16 is $\alpha \sim \pi/2$. Different curves are calculated assuming different values of the maximum PEM retardation $A$, as specified in the legend.

We note also that the angle $\alpha$ entering Eq. S17 is measured in the in-plane directions ($x'$, $y'$) of the sample reference frame, different from the reference frame ($x$, $y$) used to determine all the other angular quantities appearing in Eqs. S12-16. The rotation of the reference axes can be experimentally estimated by selecting a constant linear polarization at the entrance of the fiber probe with a known direction, by placing a rotatable linear polarizer in front of the photodetector and by looking at the maximum (or minimum) signal. Being the photodetector mounted in the SNOM microscope, the direction of the reference axes in the SNOM frame can be deduced. However, such a technique can be rather cumbersome from the experimental point of view, in particular because it requires removal and replacement of the sample. An *in-situ* procedure can be carried out instead, based on the circumstance that the calibration curves depend on the maximum retardation $A$ set for the PEM. This is shown, for instance, in Figure S3, where calibration curves for three different choices of $A$ are plotted. Hence, by repeating the same SNOM scan with different $A$-values and comparing the so-obtained AC/DC maps, it is possible to infer on the value of $\alpha$ entering Eq. S17. For the measurements shown in the paper, $\alpha$ is slightly larger than $\pi/2$. Being the nanofiber axis roughly aligned along the vertical direction of the scan, measurements are referenced to polarization directions aligned roughly along and across the nanofiber, that is, $k_1$ and $k_2$ roughly corresponds to absorption along the longitudinal and radial directions of the fiber, respectively. As a consequence, negative and positive dichroic ratios correspond to prevalent absorption of radiation polarized along or across the fiber, respectively.

## 6. Imaging of the electrospinning jet profile

For imaging the polymer jet profile a high speed camera (Photron, FASTCAM APX RS, 1024 pixel $\times$1024 pixel, 10000 frame s$^{-1}$) coupled to a stereomicroscope (Leica MZ 12.5) is used. A typical image of the measured jet shape is shown in Figure S4. The dependence of the jet radius, $a$, on the axial coordinate, $z$, is modelled by the following relation:[S8,S22]

$$\left(\frac{a_0}{a}\right) \cong \left(1 + \frac{z}{z_0}\right)^{\beta},$$
(S20)

Fitting the jet profile to Equation (S20) provides values of $z_0 = 22$ µm, and $\beta = 0.94$, using the initial radius $a_0 = 96$ µm.

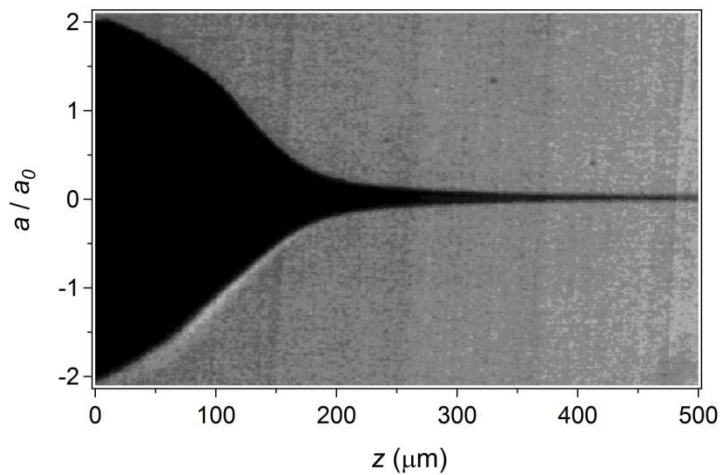

**Figure S4**. Typical image of a jet profile used to calculate the radius dependence on the axial

coordinate, $z$ (Eq. S20). $a_0 = 96$ µm.